\documentclass[a4paper,fleqn,usenatbib]{mnras}
\usepackage{graphicx}
\bibpunct{(}{)}{;}{a}{}{,}
\usepackage{txfonts}
\usepackage{longtable}
\usepackage{color}

\def\arcsec{\hbox{$^{\prime\prime}$}}

\newcommand{\xmm}{{\it XMM-Newton}}
\newcommand{\cxo}{{\it Chandra}}

\def\rahour{\hbox{\ensuremath{^{\rm h}}}}
\def\ramin{\hbox{\ensuremath{^{\rm m}}}}
\def\rasec{\hbox{\ensuremath{^{\rm s}}}}

\begin{document}
\title[]{Discovery of a radio nebula around PSR J0855--4644}
\author[Maitra et al.]{C.~Maitra$^{1}$\thanks{Contact e-mail: cmaitra@mpe.mpg.de},  S.~Roy$^{2}$, F.~Acero$^{3}$, Y.~Gupta$^{2}$\\
1.Max-Planck-Institut f\"ur extraterrestrische Physik,Giessenbachstra{\ss}e, 85748 Garching, Germany \\
2.National Centre for Radio Astrophysics, TIFR, Pune University Campus, Post Bag 3, Pune 411 007, India\\
3.Laboratoire AIM, CEA/DRF - CNRS - Universit\'{e} Paris Diderot, IRFU/DAp, CEA-Saclay, 91191 Gif-sur-Yvette, France}

\date{Accepted.....; Received .....}


\maketitle


\begin{abstract}

We report the discovery of a diffuse radio emission around PSR J0855--4644 using an upgraded {\it GMRT} ({\it uGMRT}) observation at 1.35 GHz.
The radio emission is spatially coincident with the diffuse X-ray pulsar wind nebula (PWN) seen with \xmm\  but is much larger in extent compared 
to the compact axisymmetric PWN seen with \cxo. The morphology of the emission, 
with a bright partial ring-like structure and two faint tail-like features strongly resembles a
bow shock nebula, and indicates a velocity of 100 km/s through the
ambient medium. We conclude that the emission
is most likely to be associated with the radio PWN of PSR J0855--4644. From the integrated flux density,
we estimate the energetics of the PWN.

\end{abstract}



\section{Introduction}

Young rotational powered pulsars are the powerhouses of pulsar wind nebulae (PWNe from now).
Rotational powered pulsars lose a significant part of their energy via relativistic winds which, upon interactions with the ambient medium,
produce a synchrotron powered nebula emitting from radio to beyond the X-ray bands.
The integrated energy spectrum is of synchrotron type, with a power-law having a spectral break around 10$^{13}$--10$^{15}$ Hz. 
In the radio band, the spectrum is expected to be flat, $F_{\nu} \propto \nu^{-\alpha}$, with $\alpha$ between 0 and 0.3, while 
$\alpha$ is $>1$ in the X-ray band. The steepening of the spectrum in X-rays, and hence the spectral break is most commonly associated with losses due to synchrotron cooling.
Furthermore, in the radio band the PWN luminosity traces the integrated
history of the pulsar spin-down, while in X-rays the PWN luminosity traces the current energy output of the central pulsar \citep{1984ApJ...278..630R}.
The PWN morphology provides crucial information on the properties of the outflow, the interacting ambient medium
and also the geometry of the pulsar powering it \citep{2017SSRv..207..175R}. 

PSR J0855--46444 is a young and energetic pulsar discovered by the Parkes multibeam radio survey \citep{2003MNRAS.342.1299K}. It lies in the Vela region ($l\approx265^{\circ}$,
$b\approx-1^{\circ}$) which is a complex region in the sky with many overlapping supernova remnants (SNRs) along our line of sight. Especially worth mentioning is the Vela remnant, one of the brightest 
and most extended
remnants in the sky \citep{1996MNRAS.280..252D}. Having a large angular size of $8^{\circ}$ diameter, it overlaps several SNRs such as Puppis A and RX J0852.0-4622. PSR J0855--4644 is located
on the south-eastern rim of RX J0852.0-4622 {\it aka} the Vela Jr, but is not associated with it. The measured spin period (P) of 65 ms and the period derivative
($\dot{P}$) of  $7.3  \times 10^{-15} ~s ~s^{-1}$ result in a spin-down luminosity ($\dot{E}$) of $1.1 \times 10^{36}$ erg/s (assuming a moment of inertia ($I$) of $10^{45}$ g $cm^{2}$  for standard neutron star parameters). 
The characteristic age ($\tau_{\rm c} \equiv {P}/2\dot{P}$) is estimated to be 140 kyr. 

The source was observed with \xmm, which revealed the
X-ray counterpart of the pulsar surrounded by a diffuse non-thermal extended emission which is the PWN associated with it \citep{2013A&A...551A...7A}.
Additionally, comparison of column densities provided an upper limit of 900 pc for the distance to the source. A dedicated \cxo \, observation revealed
a further compact X-ray nebula with an axisymmetric morphology resembling a double torus PWN, analogous to the Vela PWN which has an $\dot{E}$
similar to PSR J0855--4644 \citep{2017A&A...597A..75M}. However, this nearby energetic PWN has not been reported to be followed up in the radio wavelengths so far.

In this letter we report the discovery of the radio counterpart of the PWN surrounding PSR J0855--4644 using an observation with the upgraded {\it GMRT} ({ \it uGMRT}) 
\citep{gupta}. Sect. 2 presents the observations
and analysis, Sect. 3 the results, Sect. 4 the discussion on the nature of the radio emission and the inferred properties. Finally Sect. 5 presents the
summary and conclusions.
\section{Observation \& Analysis} 
Observations with the { \it uGMRT} were carried out on 30 January 2017 for 6 hours.
All 30 antennas of the array were used to obtain maximum $u-v$ coverage. The pointing centre was RA: 08\rahour55\ramin36.0\rasec \, 
and DEC:-46\rahour44\ramin13.2\rasec \,.
Observations were carried out at 1.35 GHz with a bandwidth of 200
MHz using 1024 spectral channels. 3C147 was used as primary flux density
and bandpass calibrator. The VLA calibrator 0835-451 \footnote{https://science.nrao.edu/facilities/vla/observing/callist} was observed as a
secondary calibrator. This calibrator was observed every 30 minutes. 
After calibration and editing, frequency channels were averaged by a factor of 8 to provide a channel width of 1.56 MHz in the output data. This process
significantly reduced the data volume while keeping bandwidth smearing smaller than
the synthesized beam during imaging up to half power point of the antennas.
The initial images were improved by phase-only calibration and at the last stage by an
amplitude and phase self-calibration.  The data were then used to image the
target source with a short UV cutoff of 0.7 kilo-lambda which resolved out extended structures of 
angular size $\gtrsim2.5'$ in the field. The imaging methodology used in the current work could have missed a possible large scale structure of size $\gtrsim 4'$. To verify this we made another image
with a lower UV cutoff as set by the data to 200 lambda. No other significant emission brighter than 2 $\sigma$ or $\sim$ 110 $\mu$Jy/beam (beam size $9\arcsec \times 5\arcsec$)
was seen around the emission. 
\section{Results}

\begin{figure*}
\centering
\includegraphics[scale=0.17]{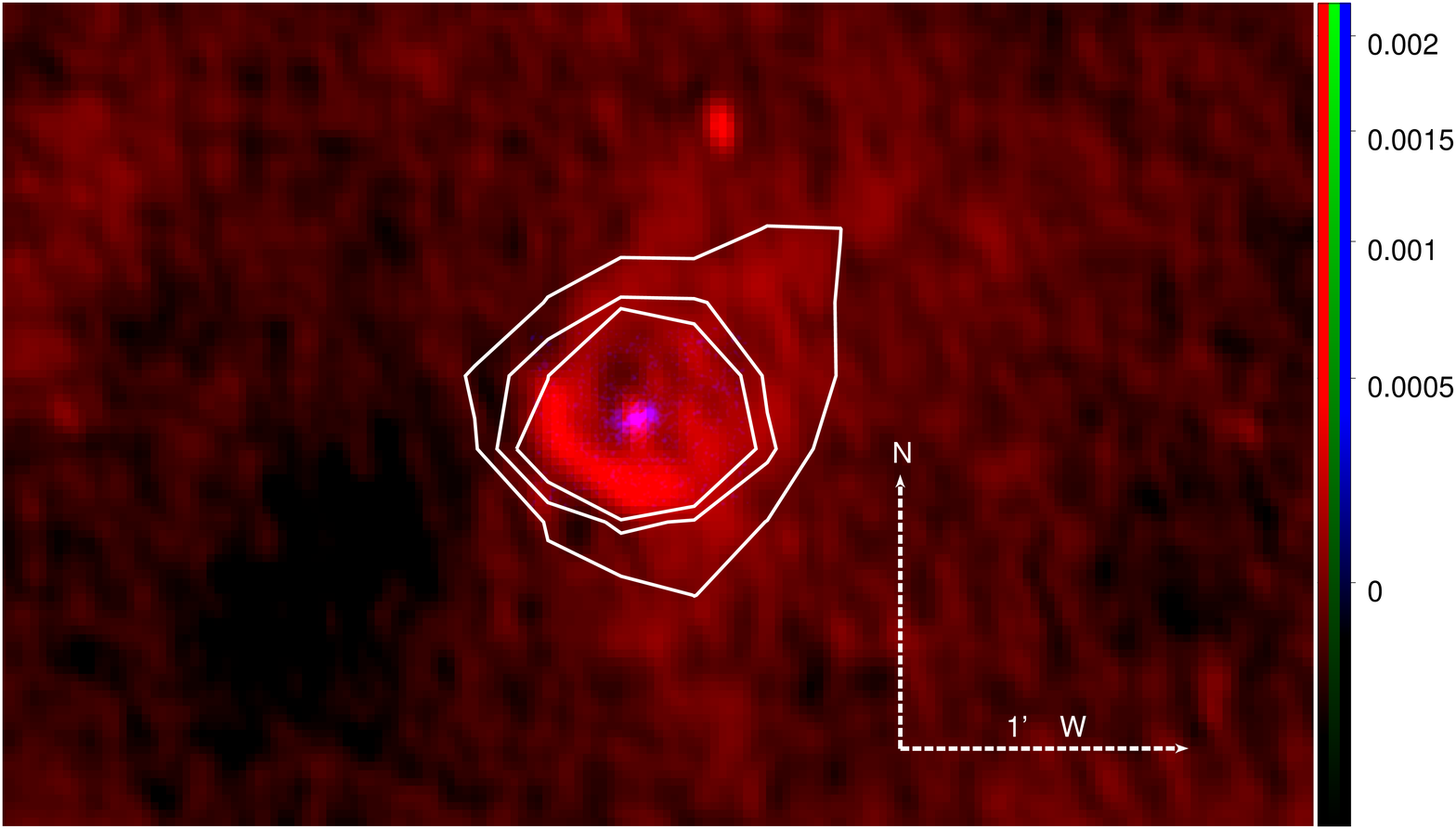}
\includegraphics[scale=0.27]{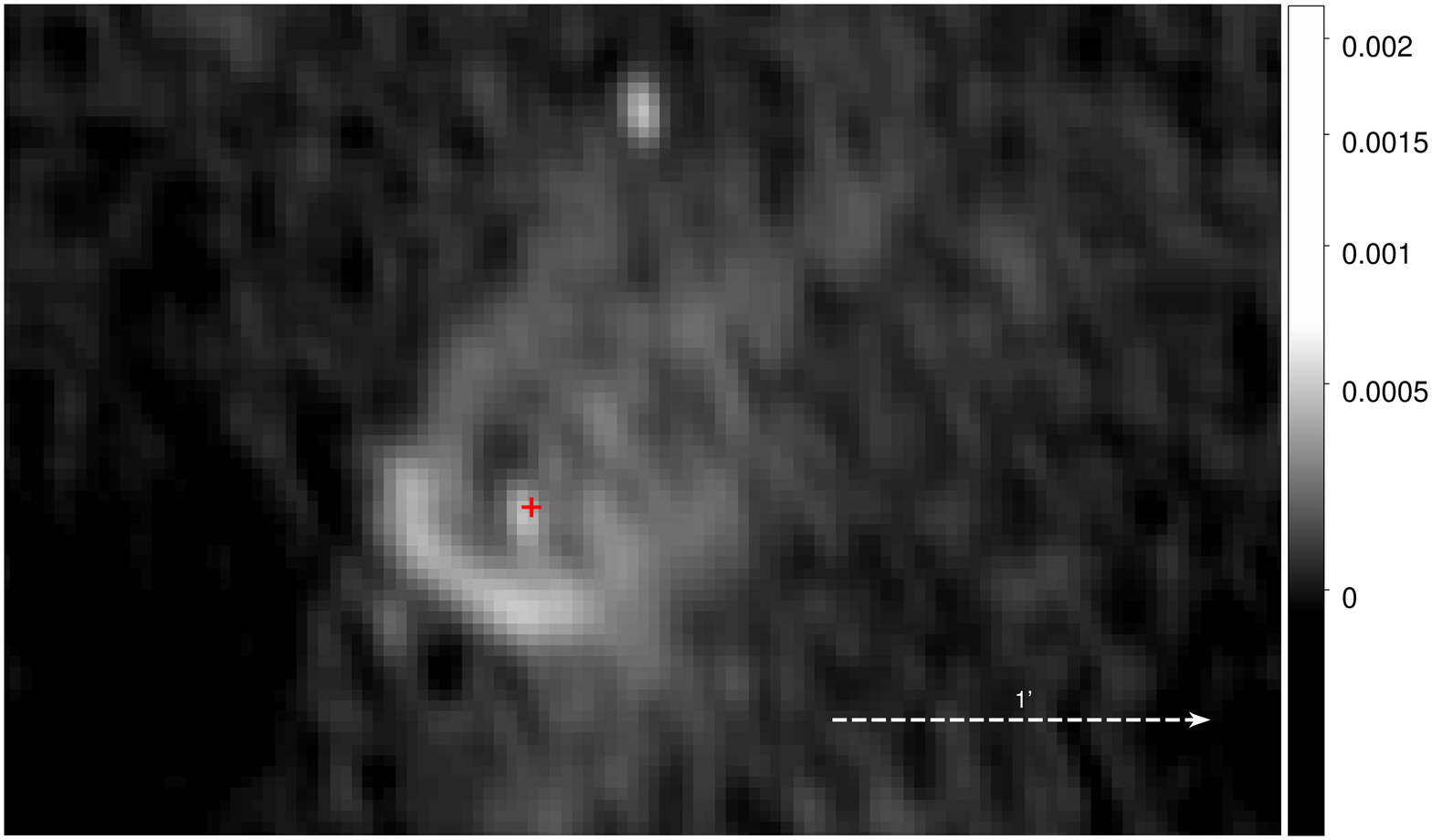}

\caption{{\it Left:}Red-Blue image of the region. Red:1.35 GHz {\it uGMRT} image of the region around PSR J0855--4644 showing diffuse 
emission around the pulsar.
Resolution of the radio image is $9\arcsec \times 5\arcsec$. 
Its major and minor axis is $9\arcsec$ and $5\arcsec$ respectively and the beam position angle is 16$^{\circ}$ (major axis orientation from North towards East).
Blue: 0.5--8 keV \cxo\, image showing the compact PWN. 
The solid white contours denote the 3, 4 and 5$\sigma$ contour around the diffuse X-ray PWN detected with \xmm~ (1.2-6 keV).
{\it Right:} A  zoomed in view showing the radio nebula and the shell structure. The red cross marks the position of the pulsar as
reported in Kramer et al. 2003.}

\label{image}
\end{figure*}

Fig. \ref{image} (right) shows the 1.35 GHz {\it uGMRT} image of the region around PSR J0855--4644 showing the diffuse structure around the pulsar. 
The integrated flux density of the radio emission is $14\pm2$ mJy, and
the significance of the emission is 7 $\sigma$. Enhanced emission is also detected at the position of the pulsar 
at a flux density of $\sim$ 0.5 mJy. This is consistent with the estimated flux of the pulsar at 1.4 GHz from \cite{2003MNRAS.342.1299K}. 
We find the position of the pulsar to be at
RA: 08\rahour55\ramin36.29\rasec \, $\pm$ 0.02\rasec \ and DEC:-46\rahour44\ramin15.29\rasec \, $\pm$ 0.42\rasec \,.
The difference in RA between the originally reported radio position and the present one is $\sim$1.5$\arcsec$.
This could arise due to the proper motion of the pulsar in the sky plane. This is however unlikely because the position of the pulsar reported
in radio \citep[][1999]{2003MNRAS.342.1299K} and in X-rays \citep[][2012]{2017A&A...597A..75M} at different epochs are consistent with each other.
In the absence of any known high resolution ($\sim$1$\arcsec$) and high sensitivity observation of the field near 1.4 GHz, 
we cannot check the positions of other background sources in the {\it GMRT} 1.35 GHz field of view. 
Field sources observed at 1.35 GHz could be systematically offset (by $\sim$1.5$\arcsec$) due to the phase change caused by ionospheric 
effects between the target field and the phase calibrator.

Fig. \ref{image} (left) shows a Red-Blue image of the region with the 1.35 GHz {\it uGMRT} image in red and the 0.5--8 keV \cxo\, image from \cite{2017A&A...597A..75M} in blue.
The white contours denote the 3, 4 and 5$\sigma$ contours of the diffuse X-ray PWN from \cite{2013A&A...551A...7A}.
The extent of the radio feature is much larger than the compact axisymmetric nebula seen with \cxo\ , and is comparable 
to the diffuse X-ray nebula seen with \xmm \,.
The central ring-like radio emission surrounding the pulsar is $\sim$ 45$\arcsec$ in diameter, and is of the same size as the inner nebula seen with \xmm. 
This corresponds to the innermost white
contour denoting the emission at 5$\sigma$ significance. A brightening is observed at the south-eastern region of the ring. 
Two extended tail-like features extend in the north-west direction making the total nebula about
$1.5\arcmin$ ($90\arcsec$) in extent. It is worth noting that \cite{2013A&A...551A...7A} reported diffuse X-ray emission up to $\sim$ $150\arcsec$. However, as can be seen from the
over-plotted contours, the 3$\sigma$ emission extends only up to $90\arcsec$. Further the X-ray emission seen with \xmm \, is contaminated with the Vela
SNR at energies $<$ 1 keV, and by the non-thermal emission from the Vela Jr at energies $>$ 1 keV. Therefore, any low significance emission should be treated
with caution.

\begin{figure}
\includegraphics[scale=0.23]{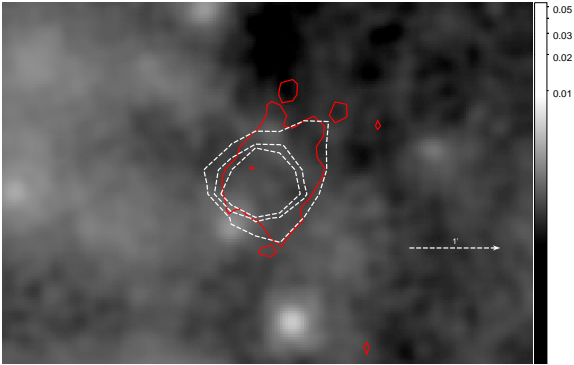}
\caption{ 22 $\mu$m {\it WISE} background subtracted image of the region around PSR J0855--4644.
Resolution of the image is $12\arcsec$. The dashed white contours denote the \xmm~ contours in Fig.~\ref{image}.
The solid red contours denote radio contour at the level of 1 mJy/beam.}
\label{image2}
\end{figure}
\section{Discussion}
At the estimated distance of 900 pc, the radio nebula detected around PSR J0855--4644 corresponds to a physical size of 0.44 pc. 
The inferences on the nature of the radio emission and some derived physical 
properties of the system are given
below.
\subsection{Nature of the radio emission}
The spatial coincidence of the enhanced radio emission with the diffuse X-ray emission
surrounding PSR J0855--4644, is a strong indication that we have 
discovered the radio counterpart of the PWN. However, in the absence of radio data at two different wavelengths, the spectral index of the source cannot
be determined which would establish the nature of the emission as non-thermal with certainty (as is expected from PWNe). In the absence of this, we have
investigated
the possibility of this emission emerging from two other competing phenomena: a SNR associated with PSR J0855--4644 and an overlapping HII region.

First we assume that the ring-like radio structure surrounding the pulsar is an associated SNR.
In this hypothesis, we can fix a lower limit for the age of the SNR to 10 kyr (the characteristic age of the PSR is 140 kyr
but this estimate is often an overestimate of the true age of the system).
At a distance of 900 pc, the physical radius is 0.1 pc. The Sedov-Taylor equation relates the radius of the forward shock 
(R$_{\rm sh}$) as a function of time (t) for a given supernova explosion energy (E), and density (n$_{0}$) as: 
\begin{equation}
 R_{\rm sh} = 5.06 ~\left(\frac{n_{0}}{1~ {\rm cm}^{-3}}\right)^{1/5} \left(\frac{E}{10^{51}~ {\rm ergs}}\right)^{1/5}
 \left(\frac{t}{10^{3}~ {\rm years}}\right)^{2/5} {\rm pc}
\end{equation}
Assuming $E=10^{51}$ ergs, n$_{0}$ is $\sim$ 10$^{10}$ cm$^{-3}$ (for $R_{\rm sh}=0.1$ pc and age t=10 kyr).
In the unlikely scenario where the pulsar distance is overestimated by a factor of 10 (i.e. $R_{\rm sh}=1$ pc), 
n$_{0}$  would be still be very high (10$^{5}$ cm$^{-3}$). Assuming a cloud of size similar to the partial 
shell-like source with a density of 10$^{10}$ cm$^{-3}$ (10$^{5}$ cm$^{-3}$), 
its associated column density ($N_{\rm H}$) would be $6 \times 10^{27}$ cm$^{-2}$ ($6 \times 10^{22}$ cm$^{-2}$).
Comparing this from the HI and $^{12}$CO data along the line of sight to the pulsar \citep[Fig. 6 of][]{2013A&A...551A...7A}, $N_{\rm H} \sim 1.5 \times 10^{22}$ cm$^{-2}$,
when integrated over all velocities. Therefore, we conclude that the association of the ring-like structure with a SNR evolving in a dense molecular
cloud is unlikely.

We also consider the possibility of the emission being thermal free-free in
nature. This region has been observed by the {\it WISE} survey at 22 $\mu$m \citep{2010AJ....140.1868W}. 
Infra-red emission at 24 $\mu$m has earlier been shown
to be highly correlated with Paschen-${\alpha}$ emission \citep{2005ApJ...633..871C} 
and is used as a proxy of thermal emission \citep{2006ApJ...638..157M} from galaxies. Emission at this band can be used to predict
the thermal emission in the radio band \citep{2013MNRAS.433.1675B}.  Dust emissivity at a very nearby wavelength of 22 $\mu$m is expected to be quite
close to its 24 $\mu$m emission and can be used as a tracer of HII region
density. The Model by \cite{2017A&A...603A.114G} indicate the dust
emission at 24 $\mu$m to be $\sim$15\% higher than what is measured at 22
$\mu$m from {\it WISE} data. From the background subtracted images, we did not detect any
emission at the position of the source from the {\it WISE} map at 22 $\mu$m (Fig.~\ref{image2}). The
measured upper limit of its flux density at 22 $\mu$m is 700 DN ({\it WISE} map
unit).  Using the conversion of DN to Jy (5.22$\times 10^{-5}$)
\footnote{http://wise2.ipac.caltech.edu/docs/release/allsky/expsup/sec2\_3f.html},
and using the expected conversion from 22 to 24 $\mu$m as described above, the
upper limit of the 24 $\mu$m flux density from the region is 0.04 Jy.  We assume
the electron temperature of any thermal gas in the region to be 10$^4$ K.
Following \cite{2013MNRAS.433.1675B}, the predicted upper limit of thermal emission at
1.35 GHz from the region is 0.3 mJy. This is almost 2 orders of magnitude
lower than what is measured at 1.35 GHz from our observation. 
As noted in \cite{2013MNRAS.433.1675B}, the conversion of 24 $\mu$m emission to radio band could be
uncertain typically by up to a factor of 2. Therefore, we conclude that the
emission from the region seen in radio cannot be free-free emission in
nature.

Having ruled out the above two scenarios, we infer that the radio emission is most likely to be associated with the PWN.
The morphology of the emission, with a bright partial ring-like structure in the south-east and two faint 
tail-like features in the north-west direction is
reminiscent of a bow shock nebula.
The direction of the westernmost tail is somewhat aligned along the spin axis of the pulsar derived in \cite{2017A&A...597A..75M} (offset by $\sim$ 10$^{\circ}$). This is in further support of its bow shock origin.
Bow shock nebulae are formed when a pulsar moves supersonically through the ambient medium.
They are most often seen either in non-recycled pulsars with ages between 10 kyr and 3 Myr, or in older recycled pulsars.
The observed spin-down luminosities ($\dot{E}$) range from $10^{33}-10^{37}$ erg/s \citep[][and references therein]{2017JPlPh..83e6301K}. 
A faint bow shock nebular structure in the radio band has also been detected in the Vela pulsar; a system analogous to PSR J0855--4644 \citep{2011ApJ...740L..26C}. 
Although the proper motion for PSR J0855--4644 is not yet known, the bow shock structure indicates supersonic motion through the local medium. In this case the termination shock 
radius ($R_{Ts}$) can be determined from the
balance between the relativistic wind of the pulsar and the ram pressure at the head of the shock. This is expressed as \citep{2004ApJ...617..480C}
 \begin{equation} 
       R_{Ts} \sim 3\times10^{16}\dot{E}^{\frac{1}{2}}_{34}n^{\frac{-1}{2}}_{1}v^{-1}_{p,100} ~cm.
      \end{equation}
  Here $n_{1}$ is the number density in units of particles cm$^{-3}$ and $v_{p,100}$ is the space velocity of the pulsar in units of 100 km/s.
Assuming the partial-ring to be the termination surface $R_{Ts}$, the distance between the pulsar and the tip
of the bow shock is
$\sim$ 20$\arcsec$ ($\sim$ 0.1 pc at a distance of 900 pc).
Given an $\dot{E}$ of $10^{36}$ erg/s and a typical ambient density of 1 cm$^{-3}$, the velocity of the pulsar is $\sim$ 100 km/s. 
Further assuming that the transverse velocity 
of the pulsar is comparable to the velocity estimated as above, a shift of $\sim$ 0.5$\arcsec$ is expected between the radio and the X-ray observation separated by 13 years.Given the uncertainty in the determination of the positions,
such a shift would not be noticeable and is consistent with our results. Future deeper and more sensitive observations separated by a large time gap will ascertain this. 
\subsection{Comparison between the radio and X-ray emission}
The difference between the size of the radio and X-ray PWN mainly depends on whether synchrotron cooling has set in.
This can often lead to a difference in size between the radio and X-ray PWN. Since the higher energy X-ray electrons cool faster, 
the number of X-ray emitting particles decrease rapidly with increasing distance from the pulsar leading to a smaller size of the X-ray nebula compared to the
the radio counterpart. For the same reason a steepening in the power-law spectral index of the synchrotron emission is also observed frequently.
The break frequency $\nu_{break}$ is a function of the age of the pulsar and the magnetic field of the nebula, {\it i.e} lower values of $\nu_{break}$ are expectedforhigher nebular magnetic fields B$_{neb}$. 
Therefore systems which have comparable sizes of the X-ray and radio nebula have either weak B$_{neb}$ or are intrinsically young
PWNe where the process of cooling has not yet
set in \citep[e.g. 3C 58, G130.7+3.1:][]{2006ARA&A..44...17G}. Furthermore, the radio and the X-ray emission are not always spatially 
correlated and even anti-correlation has been observed,
e.g., for Vela \citep{2003MNRAS.343..116D}; and G319.9-0.97 \citep{2010ApJ...712..596N}.

Comparison between the radio and X-ray emission around PSR J0855--4644 (Fig. \ref{image}) indicates that the radio nebula is comparable in size with the X-ray nebula.
The faint tail-like features of the radio emission in the north-west however extend beyond the X-ray emission. This indicates that the size of the X-ray PWN $\leq$
to the radio counterpart. The $\dot{E}$ and the morphology of the compact X-ray PWN around PSR J0855--4644 point to a Vela-like pulsar, 
generally categorized as fast-spinning pulsars with characteristic
ages 10 kyr $\lesssim$ $\tau_{\rm c}$ $\lesssim$ 100 kyr and $\dot{E}$ $\gtrsim$ $10^{36}$ erg/s \citep{2003MNRAS.342.1299K}. 
Although the characteristic age of the system is
measured to be 140 kyr, it is to be noted that this is often an overestimation of the age of the system \citep[e.g.][]{2002ApJ...567L.141M}.
Therefore PSR J0855--4644 is not a young PWN (Crab-like)
where synchrotron cooling might not have set in. This indicates that the system has a weak B$_{neb}$ which renders the cooling time-scales to be longer.

\subsection{Inferred properties of the system}
From the integrated flux density, the radio luminosity of the PWN ($L_{R}$) can be estimated assuming a typical spectral index of -0.3 between
$10^{7}$ and $10^{11}$ Hz, and using equation (4) in \cite{1997ApJ...480..364F}. The estimated $L_{R}$ is $\sim$ 
$1.3\times10^{30}$ erg/s and the radio efficiency $L_{R}$/$\dot{E}$ $\equiv$ $\eta_{R}$ is $\sim$ $1\times10^{-6}$.
Given that the flux is measured only at one frequency and not integrated over the entire frequency range of emission of the radio nebula, 
this is a lower limit. Both $L_{R}$ and $\eta_{R}$ are however
very similar to that obtained for the Vela pulsar \citep{2003MNRAS.343..116D}.

\section{Summary}
From radio observations using the {\it uGMRT} at 1.35 GHz, we have discovered diffuse radio emission surrounding the Vela-like
pulsar PSR J0855--4644, for the first time. A central ring-like radio emission feature surrounds the pulsar, which is brightened 
at the south-east region. Two tail-like features extend in the north-west direction. A faint diffuse emission fills the whole feature.
The structure corresponds to a physical size of 0.44 pc. This is much larger than the compact axisymmetric X-ray nebula (0.06 pc) seen with \cxo \,
but is similar in extent to the diffuse X-ray nebula seen with \xmm.

The spatial coincidence of the radio emission with the diffuse X-ray emission is in strong
favour of it being the radio counterpart of the PWN. We have also ruled out its origin as an overlapping SNR on a HII region along the line
of sight, given the size of the structure, its comparison with HI and $^{12}$CO data along the line of sight to the pulsar, and the {\it WISE} infrared maps.
The PWN morphology strongly resembles a bow shock nebula and the radius of the termination shock indicates a velocity of 100 km/s through the
ambient medium.

The integrated flux density at 1.35 GHz provides an estimate of $L_{R}$ $\sim$ $1.3\times10^{30}$ erg/s and $\eta_{R}$ $\sim$ $1\times10^{-6}$.
These are similar to that observed in the Vela PWN. 
\section*{Acknowledgments}
We thank the referee for providing very useful comments which helped to improve the paper.
We thank the staff of the {\it GMRT} that made these observations possible. {\it GMRT} is run by the National Centre for Radio Astrophysics of the Tata Institute of Fundamental Research.
\bibliographystyle{mnras}
\bibliography{radio}

\end{document}